\documentclass[prl,aps,twocolumn,floatfix,superscriptaddress]{revtex4-1}

\usepackage{amsmath,amssymb}
\usepackage{bm,bbm}
\usepackage{color}
\usepackage{xcolor}
\usepackage{ulem}
\usepackage[colorlinks=true,linkcolor=blue]{hyperref}
\usepackage{graphicx}
\usepackage{mathrsfs}
\usepackage{slashed}

\begin{document}

\preprint{ADP-18-23/T1071, JLAB-THY-18-2806}

\title{Flavor symmetry breaking in the $\Delta$ sea}

\author{J.~J.~Ethier}
\address{Nikhef Theory Group, Science Park 105,
	1098 XG Amsterdam, The Netherlands}
\author{W.~Melnitchouk}
\address{Jefferson Lab, 12000 Jefferson Avenue,
	Newport News, Virginia 23606, USA}
\author{Fernanda~Steffens}
\address{Instit\"ut f\"ur Strahlen- und Kernphysik,
	Universit\"at Bonn, Nussallee 14-16,
	53115 Bonn, Germany}
\author{A.~W.~Thomas}
\address{CSSM and ARC Centre of Excellence for Particle Physics
	at the Terascale, Department of Physics,
	University of Adelaide SA 5005, Australia}

\begin{abstract}
The discovery of a sizeable asymmetry in the $\bar u$ and $\bar d$
distributions in the proton was one of the more consequential
experimental findings in hadron physics last century.
Although widely believed to be related to the fundamental role
of chiral symmetry in QCD, a definitive verification of this
hypothesis has remained elusive.
We propose a novel test of the role of chiral symmetry in
generating the sea flavor asymmetry by comparing the $\bar d-\bar u$
content in the proton with that in the $\Delta^+$ baryon, where a
significant enhancement is expected around the opening of the $N \pi$
decay channel.
Recent developments in lattice QCD suggest a promising way
to test this prediction in the near future.
\end{abstract}

\date{\today}
\maketitle

As a result of considerable theoretical and experimental effort
we now know that the sea of quark-antiquark pairs in the nucleon
is far more complex than originally envisaged on the basis of
simple quark models or perturbative QCD.
The first major surprise was the confirmation in the early 1990s
of an integrated excess of $\bar d$ over $\bar u$ antiquarks in
the proton~\cite{Amaudruz:1991at}, leading to a violation of the
Gottfried sum rule~\cite{Gottfried:1967kk}.
Almost a decade earlier, as a by-product of a study of the excess
of non-strange over strange sea quarks predicted within the
cloudy bag model~\cite{Thomas:1982kv, Theberge:1980ye},
it had been shown that the application of chiral symmetry
to the structure of the nucleon naturally led to a surplus
of $\bar d$ over $\bar u$~\cite{Thomas:1983fh}.

Once the experimental result was announced, a number of calculations
confirmed that the pion cloud picture could indeed explain it
quantitatively~\cite{Melnitchouk:1991ui, Kumano:1991em, Henley:1990kw,
Speth:1996pz}.
Furthermore, careful study of the nonanalytic behavior of the sea
quarks as a function of quark mass established that the pion cloud
contribution was an essential feature of spontaneous symmetry breaking
in QCD~\cite{Thomas:2000ny, Detmold2001, Chen:2001et, Arndt2002,
Chen:2001pva}.
Studies of the sea using Drell-Yan lepton-pair
production~\cite{Baldit:1994jk} in $p \bar p$ collisions at Fermilab
suggested an unexpected change of sign in $\bar d - \bar u$
at parton momentum fractions $x$ around 0.3~\cite{Hawker:1998ty},
which is difficult to accommodate naturally within a meson cloud
framework~\cite{Melnitchouk:1998rv}.
While we await the results of the follow-up SeaQuest
experiment~\cite{SeaQuest}, designed to explore the asymmetry to
larger $x$, it is imperative to obtain independent confirmation
of the physical mechanism.

In this Letter we suggest that a comparison of the $\bar d-\bar u$
asymmetry in the $\Delta^+$ baryon with that in the proton
provides an outstanding opportunity for such a confirmation.
To understand why, we recall that the dominant meson-baryon
component of the proton wave function arises from quantum
fluctuation $p \to n \pi^+$.
As the $\pi^+$ contains only a valence $\bar d$ antiquark,
one naturally expects $\bar d > \bar u$ in the proton.
The process $p \to p \pi^0$, which is suppressed by a factor
of two by isospin couplings, produces equal numbers of $\bar d$
and $\bar u$ and therefore does not affect the asymmetry.
While the process $N \to \Delta \pi$ acts to reduce the asymmetry,
it is suppressed relative to the dominant process $N \to N \pi$.

For the $\Delta^+$ baryon, the processes $\Delta \to \Delta \pi$
and $\Delta \to N \pi$ both favor $\pi^+$ production, and hence
also produce an excess of $\bar d$ over $\bar u$.
The key difference, however, is that because the $\Delta$ decay
to $N \pi$ is favored energetically, it experiences a significant
kinematical enhancement as a function of the pion mass, $m_\pi$,
as it approaches the $\Delta - N$ mass difference and the decay
channel opens up.

In parallel developments, recent progress in the calculation of PDFs
in lattice QCD suggests a realistic means to check the prediction.
In particular, lattice QCD measurement of the spatial correlation
function of quarks within a fast moving hadron could be used
\cite{Ji:2013dva}, after Fourier transformation and renormalization,
to obtain a quasi-PDF~\cite{Alexandrou:2017huk, Chen:2017mzz,
Green:2017xeu}, which through a further matching procedure
\cite{Xiong:2013bka, Alexandrou:2015rja, Izubuchi:2018srq}
can directly yield the desired light-cone PDF over the range
$x \in (-1,+1)$.
Previous attempts to extract antiquark distributions from lattice
QCD were impaired by the difficulty of disentangling the $q$ and
$\bar q$ content using only the first two or three moments from
calculations of matrix elements of local twist-two
operators~\cite{Detmold:2003rq}.
In constrast, in the quasi-PDF approach one can use the crossing
symmetry relation, $\bar q(x) = -q(-x)$, to extract directly the
$x$ dependence of the $\bar q$ PDFs.
Early exploratory studies of quasi-PDFs~\cite{Alexandrou:2015rja,
Lin:2014zya, Chen:2016utp, Alexandrou:2016jqi} indeed suggested an
asymmetric sea, even though renormalization was not yet available,
and the computations were performed at large pion masses.

Recently, however, simulations at the physical pion mass,
including a sophisticated treatment of renormalization,
have shown a promising degree of agreement with empirical
distributions~\cite{Alexandrou:2018pbm, Chen:2018xof}.
Nonetheless, a number of systematics, such as discretization
and volume effects, as well as difficulties in dealing with
high momentum hadrons on the lattice, have to be addressed
before quantitative comparisons with phenomenology are possible.
In this spirit, a measurement of the distribution $u-d$ in the
$\Delta^+$ would be of enormous interest, especially if the
difference between the $u-d$ shapes in the $\Delta^+$ and proton
is sufficiently large compared to the present computational
uncertainties.

\begin{figure}[t]
\centering
\includegraphics[width=0.45\textwidth]{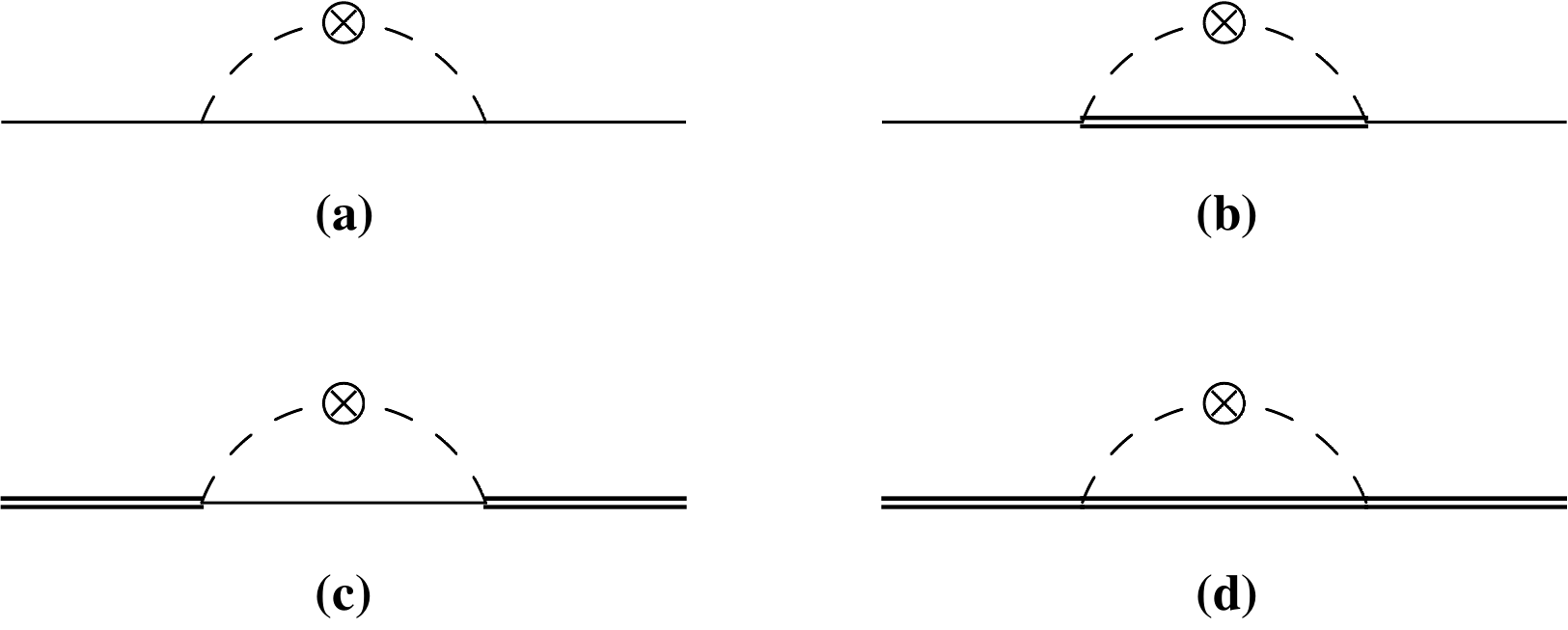}
\caption{Pion loop diagrams contributing to the $\bar d-\bar u$
	PDFs in the nucleon (solid lines) and
	$\Delta$ (double solid lines) from the processes
	(a) $N \to N \pi$, (b) $N \to \Delta \pi$,
	(c) $\Delta \to N \pi$ and (d) $\Delta \to \Delta \pi$,
	with the $\otimes$ representing the insertion of a
	nonlocal current operator.}
\label{fig:ND}
\end{figure}

Within a chiral effective theory framework, the asymmetry between
the $\bar d$ and $\bar u$ PDFs in a baryon $B$ ($B = N$ or $\Delta$)
arises through a convolution of the valence antiquark distribution in
the pion, $\bar q_v^\pi$, and the corresponding light-cone momentum
distribution, $f_{B \to B' \pi}$, of pions in $B$ with a spectator
baryon $B'$~\cite{Chen:2001pva, Burkardt:2012hk, Ji:2013bca,
Salamu:2014pka, Wang2016}.
The coupling of the external probe to the pion field in the
effective theory arises through the rainbow diagrams illustrated
in Fig.~\ref{fig:ND}, as well as via bubble diagrams in which
the pion loop couples to the baryon $B$ via a Weinberg-Tomazawa
four-point interaction~\cite{Chen:2001pva, Arndt2002,
Burkardt:2012hk, Ji:2013bca, Salamu:2014pka, Wang2016}.
The latter involve pions with zero momentum fractions $y$,
and are localized to $x=0$.
Since lattice QCD simulations cannot access PDFs at $x=0$,
the bubble diagrams will not be relevant here.

Moreover, the rainbow diagrams themselves receive zero mode
contributions~\cite{Chen:2001pva, Burkardt:2012hk},
in addition to the usual on-shell terms at $x > 0$.
Off-shell and Kroll-Ruderman terms contribute to the quark
distributions through coupling to the intermediate state
baryon $B'$~\cite{Ji:2013bca, Wang2016}.
In the following the distributions $f_{B \to B' \pi}$
(which are also referred to as chiral splitting functions)
will denote only the on-shell components of the rainbow
diagrams at $y > 0$.
The dominant contributions to the antiquark asymmetry in the
proton and $\Delta^+$ are then given by
\begin{eqnarray}
\big( \bar d - \bar u \big)^p\!(x)
&=& 2 \Big[
	\big( f_{N \to N \pi} - f_{N \to \Delta \pi} \big)
	\otimes \bar q_v^\pi
      \Big](x)
\label{eq:asymmp}
\end{eqnarray}
and
\begin{eqnarray}
\big( \bar d - \bar u \big)^{\Delta^+}\!\!(x)
&=& \Big[
      \big( f_{\Delta \to N \pi} + 2 f_{\Delta \to \Delta \pi} \big)
      \otimes \bar q_v^\pi
    \Big](x)
\label{eq:asymmD}
\end{eqnarray}
where the symbol ``$\otimes$'' denotes the convolution operator
$[f \otimes g](x) \equiv \int_x^1 (dy/y) f(y)\, g(x/y)$.

For a proton target, the $N \to N \pi$ splitting function
for Fig.~\ref{fig:ND}(a) is given by the familiar
expression~\cite{Thomas:1983fh, Burkardt:2012hk, Drell:1969wd}
\begin{eqnarray}
f_{N \to N \pi}(y)
&=& \frac{g_A^2 M^2}{(4 \pi f_{\pi})^2}\,
    \int\!dk_\perp^2\,
    \frac{y\, (k_\perp^2 + y^2 M^2)}{(1-y)^2 D_{N N}^2},
\label{eq.fNNpi}
\end{eqnarray}
where $g_A$ is the nucleon axial charge, $f_\pi$ is the pion decay
constant, $M$ is the nucleon mass, and $k_\perp$ is the transverse
momentum of the pion.
The function $D_{N N}$ is the pion virtuality $k^2-m_\pi^2$, which
in general depends on the initial and final state baryon masses,
$M_B$ and $M_{B'}$, respectively,
\begin{eqnarray}
\hspace*{-0.5cm} D_{B B'}
&=& -\frac{k_\perp^2 - y(1-y) M_B^2 + y M_{B'}^2 + (1-y) m_\pi^2}{1-y}.
\end{eqnarray}

For the corresponding process $N \to \Delta \pi$ in
Fig.~\ref{fig:ND}(b), the splitting function is given
by~\cite{Salamu:2014pka}
\begin{eqnarray}
f_{N \to \Delta \pi}(y)
&=& \frac{g_A^2}{25 M_\Delta^2 (4 \pi f_\pi)^2}
    \int\!dk_\perp^2\,
    \frac{y\, (\overline{M}^2 - m_\pi^2)}{1-y}
    \hspace*{0.3cm}				\nonumber\\
& & \hspace*{-2cm} \times
    \Bigg[
      \frac{(\overline{M}^2 - m_\pi^2) (\Delta^2 - m_\pi^2)}
	   {D_{N \Delta}^2}
    - \frac{\overline{M}^2 - 3 m_\pi^2 + 2 \Delta^2}
	   {D_{N \Delta}}
    \Bigg],
\label{eq.fNDpi}
\end{eqnarray}
where $M_\Delta$ is the $\Delta$ mass, and we have defined
$\overline{M} \equiv M + M_\Delta$ and
$\Delta \equiv M_\Delta - M$.
In the chiral limit, moments of the splitting functions can be
expanded in power series in $m_\pi$, with the leading nonanalytic
terms in the expansion, which depend only on the long-distance
properties of pion loops, being model independent~\cite{Thomas:2000fa}.
For the $N \to N \pi$ distribution one finds the characteristic
leading order (LO) $\sim m_\pi^2 \log m_\pi^2$ nonanalytic
behavior~\cite{Thomas:2000ny, Detmold2001, Chen:2001et, Arndt2002,
Chen:2001pva}.
Moments of the $N \to \Delta \pi$ splitting function,
in contrast, display the next-to-leading order (NLO) behavior
$\sim m_\pi^4 \log m_\pi^2$ for $m_\pi \to 0$
\cite{Thomas:2000ny, Arndt2002, Salamu:2014pka, Wang2016}.

In the case of a $\Delta$ baryon initial state, the LO contribution
is given by
\begin{eqnarray}
f_{\Delta \to \Delta \pi}(y)
&=& \frac{g_A^2}{50 M_\Delta^2 (4 \pi f_\pi)^2}
    \int dk^2_\perp \frac{y}{1-y} \hspace*{2cm}		\nonumber\\
& & \hspace*{0.3cm} \times
    \Bigg[
      \frac{m_\pi^2
	    \big[m_\pi^2 (2M_\Delta^2-m_\pi^2) - 10 M_\Delta^4\big]}
	   {D_{\Delta\Delta}^2}				\nonumber\\
& & \hspace*{1cm}
   +\ \frac{m_\pi^2 (4 M_\Delta^2 - 3 m_\pi^2) - 10 M_\Delta^4}
	   {D_{\Delta\Delta}}
    \Bigg],
\label{eq.fDDpi}
\end{eqnarray}
while the NLO distribution is
\begin{eqnarray}
f_{\Delta \to N \pi}(y)
&=& \frac{g_A^2}{50 M_\Delta^2 (4 \pi f_\pi)^2}
    \int dk^2_\perp
    \frac{y (\overline{M}^2 - m_\pi^2)}{(1-y)}
    \hspace*{0.3cm}				\nonumber\\
& & \hspace*{-1.8cm} \times
    \Bigg[
      \frac{(\overline{M}^2 - m_\pi^2)(\Delta^2 - m_\pi^2)}
	   {D_{\Delta N}^2}
    - \frac{\overline{M}^2 - 3 m_\pi^2 + 2 \Delta^2}
	   {D_{\Delta N}}
    \Bigg].
\label{eq.fDNpi}
\end{eqnarray}
In Eqs.~(\ref{eq.fNNpi})--(\ref{eq.fDNpi}) SU(6) symmetry and the
Goldberger-Treiman relation have been used to write the $\pi NN$,
$\pi N \Delta$ and $\pi \Delta \Delta$ couplings in terms of the
common ratio $g_A/f_\pi$.

The splitting functions (\ref{eq.fNNpi})--(\ref{eq.fDNpi}) are
ultraviolet divergent and therefore need to be regularized.
In the literature various regularization schemes have been advocated,
including transverse momentum cutoff, Pauli-Villars and dimensional
regularization (DR), as well as form factors or finite-range
regulators~\cite{Speth:1996pz, Young2003, Ji:2013bca, Salamu:2014pka,
Wang2016}.
The latter take into account the finite size of hadrons, while schemes
such as DR are generally more suitable for theories that treat hadrons
as pointlike.  The advantage of DR is that specific power counting
schemes can be preserved in chiral perturbation theory expansions,
whereas finite-range regulators effectively resum terms in the chiral
series.  In practice this allows for better convergence in $m_\pi$
in regions where the usual power counting schemes would not otherwise
be applicable~\cite{Young2003, Hall2010}.

\begin{figure}[t]
\centering
\hspace*{-0.2cm}\includegraphics[width=0.5\textwidth]{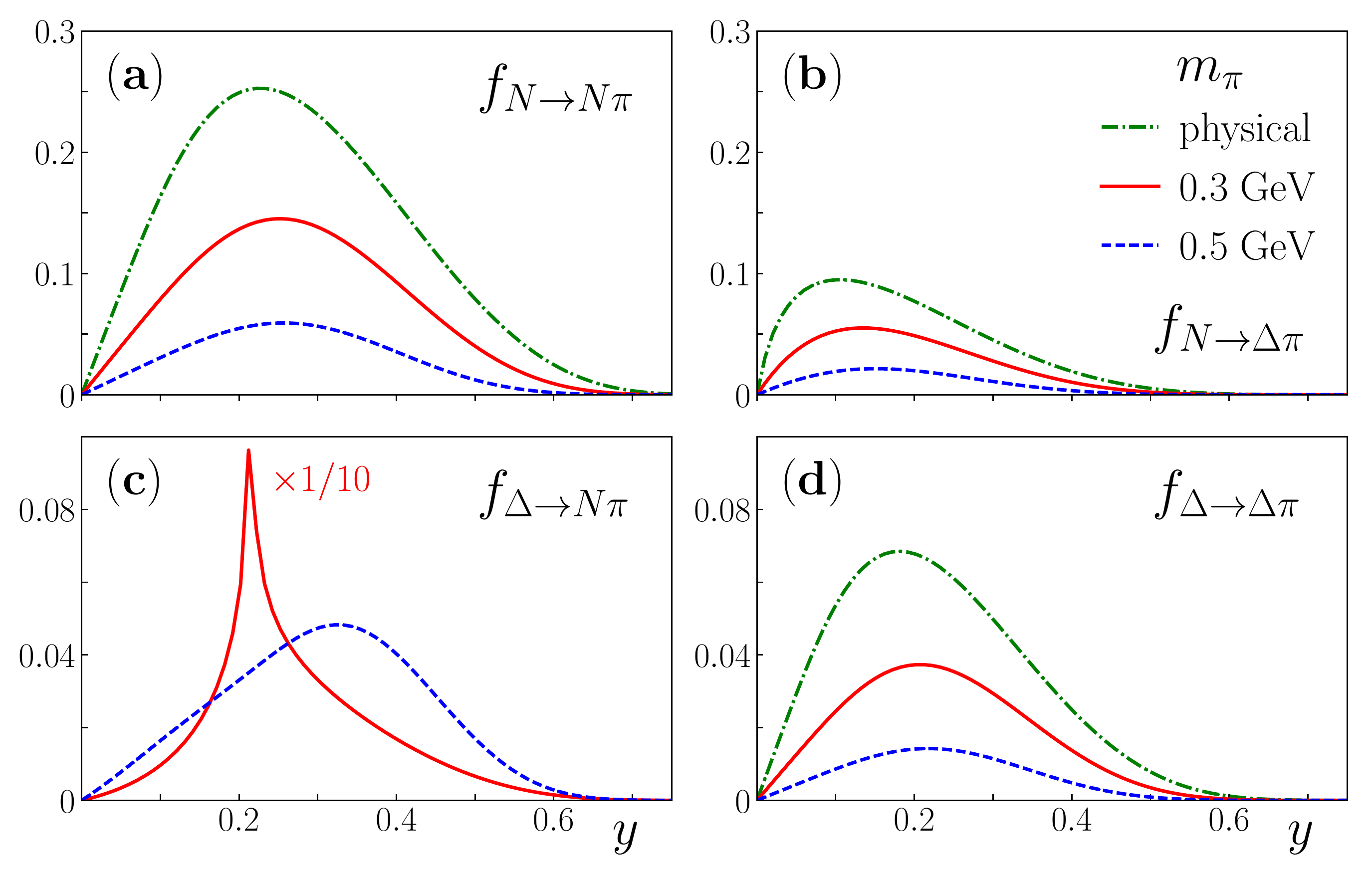}
\caption{Chiral splitting functions versus $y$ for the
	(a) $N \to N \pi$,
	(b) $N \to \Delta \pi$,
	(c) $\Delta \to N \pi$ and
	(d) $\Delta \to \Delta \pi$ transitions
	at the physical pion mass (red solid curves),
	$m_\pi=0.3$~GeV	(blue dashed curves), and
	$m_\pi=0.5$~GeV	(green dot-dashed curves),
	using the exponential regulator with cutoff mass
	$\Lambda=0.87$~GeV~\cite{Barry2018}.
	Note that the $\Delta \to N \pi$ function for
	$m_\pi=0.3$~GeV	is scaled by a factor 1/10.}
\label{fig:fy}
\end{figure}

Following the phenomenological analyses~\cite{McKenney2016, Barry2018}
of leading neutron deep-inelastic production data and other observables
sensitive to chiral loops, we consider several forms for the regulator,
including a $k_\perp$ cutoff, $k^2$-dependent exponential and monopole
form factors, and a Regge theory motivated form~\cite{Kopeliovich2012}.
In Fig.~\ref{fig:fy} we illustrate the four splitting functions
(\ref{eq.fNNpi})--(\ref{eq.fDNpi}) for a number of values of
$m_\pi$ relevant for lattice QCD simulations, for the case of the
exponential form factor with cutoff mass $\Lambda = 0.87$~GeV.
This value was obtained from the recent JAM global pion PDF
analysis~\cite{Barry2018} including constraints on $\bar d-\bar u$
in the proton from $pp$ and $pd$ Drell-Yan data~\cite{Hawker:1998ty}.
We take the same cutoff value for the $N \Delta \pi$ and
$\Delta \Delta \pi$ couplings, and assume it to be independent
of $m_\pi$ (an assumptions which is expected to break down at
large $m_\pi$).
The nucleon and $\Delta$ masses do have $m_\pi$ dependence, on the
other hand, and for these we take the approximate relations
  $M \approx M^{(0)} + m_\pi$ and
  $M_\Delta \approx M_\Delta^{(0)} + m_\pi$,
with the chiral limit values
  $M^{(0)} = 0.8$~GeV and
  $M_\Delta^{(0)} = 1.1$~GeV~\cite{WalkerLoud:2008pj}.

For the case of the nucleon initial state, the dominance of the LO over
the NLO contribution is obvious from Fig.~\ref{fig:fy}(a) and (b).
The reason is not only the smaller coupling but also the cost in energy
to convert the nucleon into a $\Delta$.
On the other hand, for a $\Delta$ initial state the enhancement
associated with the exothermic nature of the NLO $\Delta \to N \pi$
process means that it is larger than the LO $N \to N \pi$ contribution
at all pion masses, and is also larger than the $N \to \Delta \pi$
function.
At $m_\pi = 0.3$~GeV the most promiment feature in the
$\Delta \to N \pi$ splitting function in Fig.~\ref{fig:fy}(b)
is the large cusp at $y \approx 0.2$, which indicates the opening
of the octet decay channel at $m_\pi = \Delta$ (in the present
analysis we take the mass difference $\Delta \approx 0.3$~GeV
independent of $m_\pi$).
Below this threshold the $\Delta \to N \pi$ function is complex,
and is not shown in Fig.~\ref{fig:fy}(c) at the physical pion mass.
Compared to excited baryon masses, which are found to be relatively
smooth functions of $m_\pi$ across the pion decay threshold
\cite{Wright2000, Young2003}, the additional pion propagator
in the splitting function enhances the singularity at
$m_\pi \approx \Delta$ to produce the observed spike.
A similar behavior would also be expected for electroweak form
factors, and indeed was observed in the calculation of pion loop
corrections to the $\Delta$ magnetic moments ~\cite{Cloet2003}.

\begin{figure}[t]
\centering
\hspace*{-0.2cm}\includegraphics[width=0.5\textwidth]{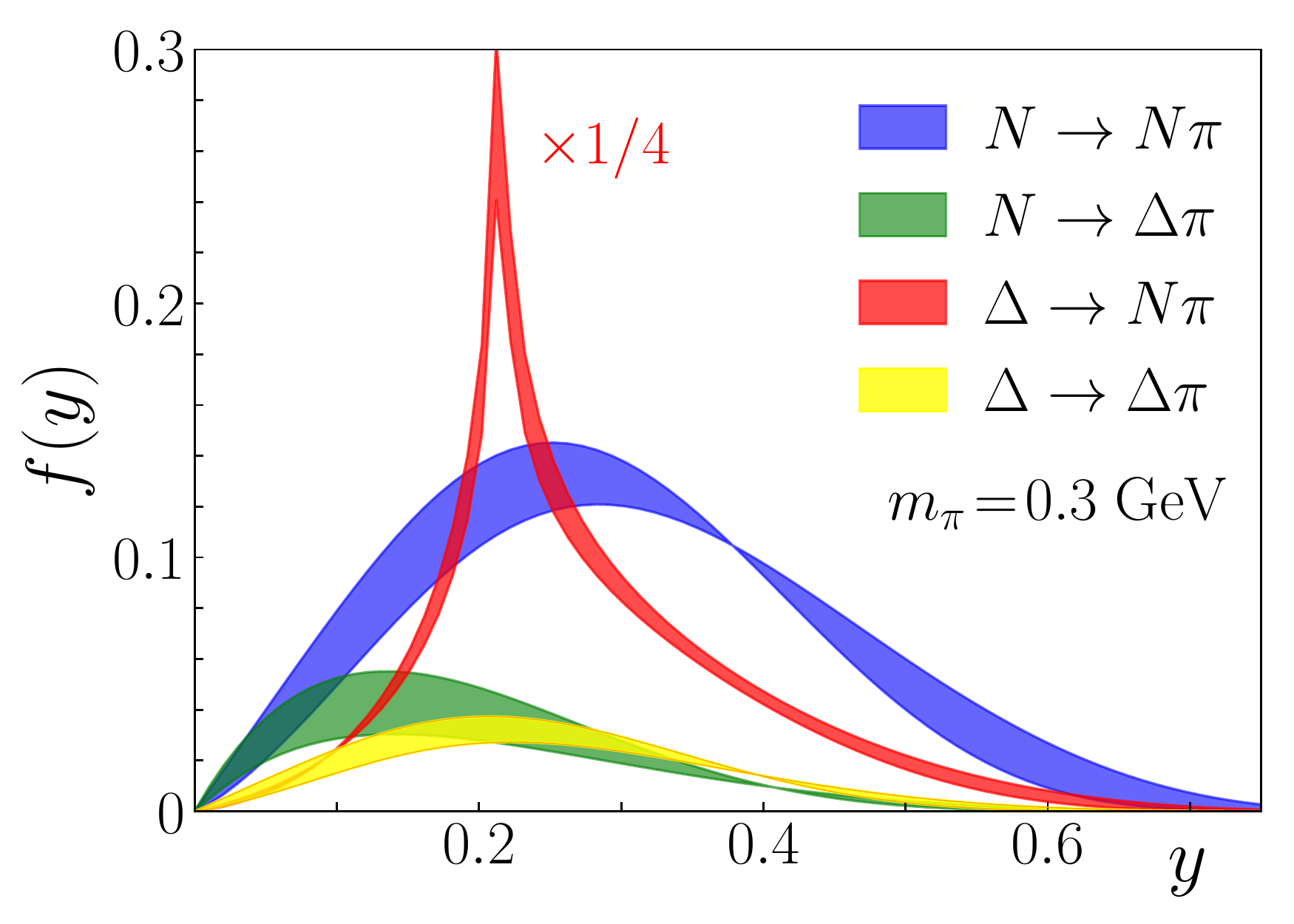}
\caption{Chiral splitting functions at $m_\pi=0.3$~GeV,	with the
	shaded bands representing the spread using the exponential
	and Regge form factor regulators~\cite{Barry2018}.
	Note that the $\Delta \to N \pi$ function is scaled
	by a factor 1/4.}
\label{fig:fy2}
\end{figure}

To make a more direct comparison of the four processes, in
Fig.~\ref{fig:fy2} we compare the splitting functions at a fixed
$m_\pi = 0.3$~GeV, at which the differences between the nucleon
and $\Delta$ splitting functions are most dramatic.
To explore the dependence of the results on the choice of regulator,
we compare the results for the exponential form factor with those
using the Regge form with cutoff $\Lambda=1.43$~GeV, also from
the JAM global PDF analysis~\cite{Barry2018}.
The shaded bands in Fig.~\ref{fig:fy2} represent the spread
between the two calculations.
As already indicated in Fig.~\ref{fig:fy}, at this $m_\pi$ value
the $\Delta \to N \pi$ channel dominates, and the presence of the
prominent cusp at $y \approx 0.2$ is independent of the choice
of regulator.
The contributions to the $N$ and $\Delta$ splitting functions
from the processes with $\Delta \pi$ intermediate states are
significantly smaller than those for the $N \pi$ channels,
regardless of the regulator form.

\begin{figure}[t]
\centering
\hspace*{-0.2cm}\includegraphics[width=0.5\textwidth]{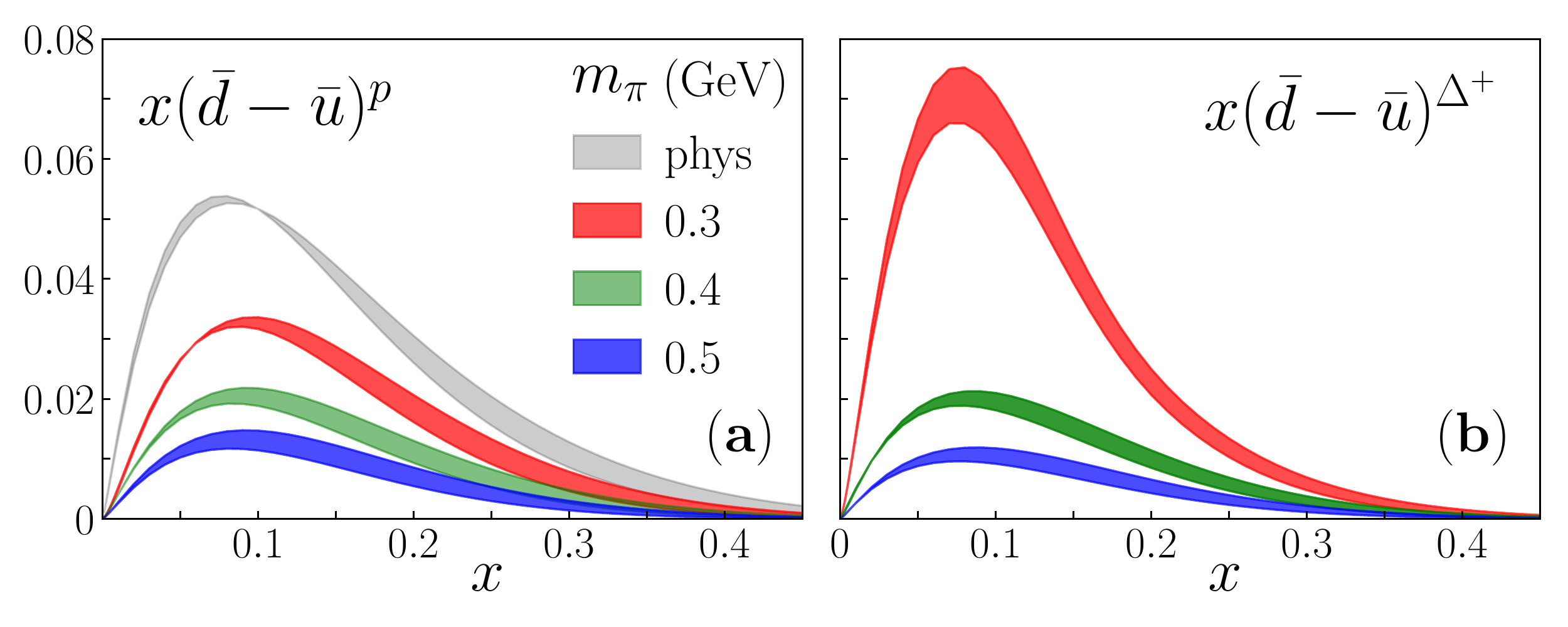}
\caption{Predicted $x$ dependence of the $x(\bar d - \bar u)$
	asymmetry in (a) the proton and (b) $\Delta^+$ baryon,
	for various pion masses:
	physical $m_\pi$ (gray band),
	$m_\pi = 0.3$~GeV (red),
	0.4~GeV (green), and
	0.5~GeV (blue).
	The shaded bands represent the model dependence
	from the choice of regulator for the splitting
	function~\cite{Barry2018}.}
\label{fig:dbar_ubar_mpi}
\end{figure}

To obtain the $x$ dependence of the $\bar d-\bar u$ distributions,
the splitting functions in Figs.~\ref{fig:fy} and \ref{fig:fy2}
need to be convoluted with the pion PDF.  While pion valence PDF
is relatively well determined from global next-to-leading-order
analyses of Drell-Yan and other high energy scattering
data~\cite{Barry2018, SMRS, Aicher2010}, its dependence on
$m_\pi$ is less well understood.
In the absence of direct lattice calculations of $\bar q_v^\pi$,
Detmold {\it et al.}~\cite{Detmold:2003tm} used the several low
PDF moments from lattice QCD simulations of pion twist-two matrix
elements to reconstruct the $x$ dependence over a range of pion
masses from the chiral limit to $m_\pi = 1$~GeV, at a scale
$Q^2 \sim 5$~GeV$^2$ set by the lattice spacing~\cite{Best1997}.

Using these inputs, in Fig.~\ref{fig:dbar_ubar_mpi} we show the
resulting $\bar d - \bar u$ asymmetry in the proton and $\Delta^+$
for several $m_\pi$ values ranging from the physical value
(for the proton only) to $m_\pi = 0.5$~GeV.
The bands in Fig.~\ref{fig:dbar_ubar_mpi} represent uncertainties
from the choice of ultraviolet regulator, corresponding to the
spread in the splitting functions shown in Fig.~\ref{fig:fy2}.
While the magnitude of the asymmetry in the proton and $\Delta^+$
are similar for large values of $m_\pi \gtrsim 0.4$~GeV$^2$,
the enhancement due to the opening of the decay channel at
$m_\pi = \Delta$ renders the asymmetry in the $\Delta^+$ twice
as large near the peak in $x(\bar d-\bar u)$ at $x \approx 0.1$.

\begin{figure}[t]
\centering
\hspace*{-0.2cm}\includegraphics[width=0.5\textwidth]{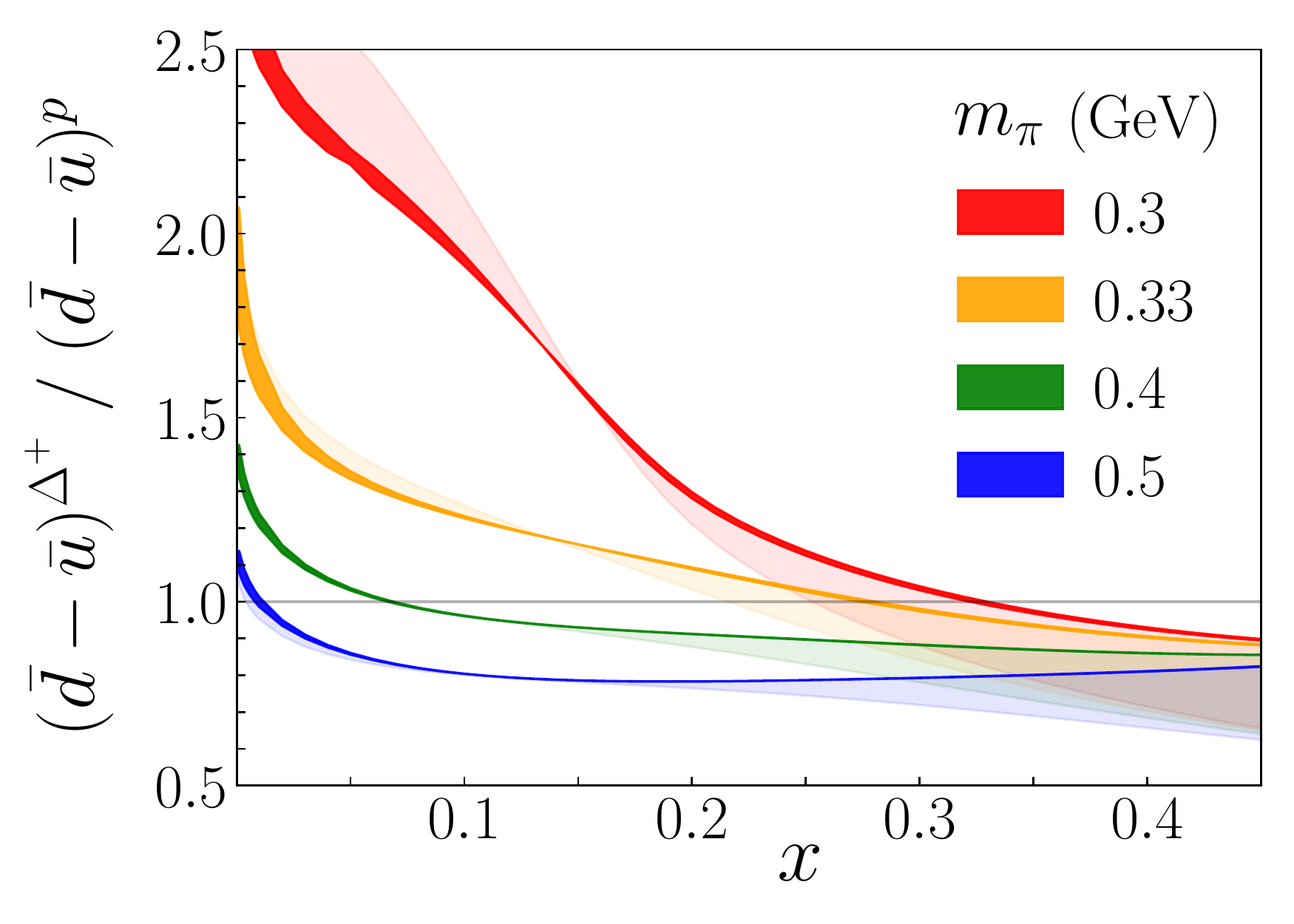}
\caption{Ratio of the $\bar d - \bar u$ asymmetry in the $\Delta^+$ to
	that in the proton, for $m_\pi = 0.3$ (red), 0.33 (orange),
	0.4 (green) and	0.5~GeV (blue bands).
	The darker bands represent the uncertainty on the pion PDF
	$\bar q_v^\pi$, while the lighter bands represent the
	dependence on the choice of regulator.}
\label{fig:dbar_ubar_latt}
\end{figure}

The model dependence is expected to cancel to some extent in the
ratio of the $\bar d-\bar u$ asymmetries in the $\Delta^+$ and $p$,
as illustrated in Fig.~\ref{fig:dbar_ubar_latt}, where the lighter
bands show the effect of the variation of the splitting functions
in Fig.~\ref{fig:fy2} for the different regulators.
To highlight the strong enhancement of the $\Delta^+$ asymmetry
as one approaches the $N \pi$ threshold, we compute the ratio at
$m_\pi=0.3$ and 0.33~GeV (at which $(m_\pi-\Delta)/m_\pi \approx 10\%$),
in addition to the 0.4 and 0.5~GeV values.

The variation with $m_\pi$ is dramatic at $x \approx 0.1$, where
the ratio goes from being $\approx 80\%$ at $m_\pi = 0.5$~GeV to
$\gtrsim 200\%$ just above the threshold at $m_\pi = 0.3$~GeV.
At larger $x$ values, $x \gtrsim 0.25$, the ratios are close to
unity for all the $m_\pi$ values considered, albeit with larger
uncertainties.  In this region the asymmetries are very small,
however, and will in practice be difficult to extract from lattice
or experiment.

The dependence of the asymmetry ratio on the input pion valence
PDF is also relatively weak, as the darker bands in
Fig.~\ref{fig:dbar_ubar_latt} illustrate.
The bands represent the difference between the results using the
splitting functions computed with the exponential regulator and
the $m_\pi$ dependent pion PDF from Ref.~\cite{Detmold:2003tm}
with those using a fixed $\bar q_v^\pi$ PDF at the physical
pion mass.
Since the same pion PDF enters both the $\Delta^+$ and proton
convolutions in the numerator and denominator for any $m_\pi$,
the dependence on $\bar q_v^\pi$ largely cancels, as expected.

The predicted large enhancement of the $\bar d - \bar u$ asymmetry
in the $\Delta^+$ can be tested in lattice QCD simulations at pion
masses just above the $N \pi$ threshold where the $\Delta$ is stable.
In particular, the ETM Collaboration plans to calculate the $u-d$
quasi-PDF in the $\Delta$~\cite{LatticeProposal} using the Iwasaki
improved gluon action and the twisted mass fermion action with clover
improvement~\cite{Alexandrou:2018pbm}. The ensembles to be used in
these simulations should allow access to $m_\pi$ values at which
$(m_\pi - \Delta)/m_\pi \approx 3\%$~\cite{Alexandrou:2015rja},
which could provide a striking confirmation of the role of chiral
symmetry and the pion cloud in the generation of a nonperturbative
sea in baryons.

This work was supported by the University of Adelaide and by the
  Australian Research Council through the ARC Centre of Excellence
for Particle Physics at the Terascale (CE110001104) and Discovery
Project DP150103164, and
  the U.S. Department of Energy (DOE) Contract No.~DE-AC05-06OR23177,
under which Jefferson Science Associates, LLC operates Jefferson Lab.
F. S. was funded by DFG Project No. 392578569.


\end{document}